\documentstyle[11pt,a4,
amssymb]{article}
\input{epsf}
\newcommand{\beq}[1]{\begin{equation}\label{#1}}
\newcommand{\eeq}{\end{equation}}
\newcommand{\bear}[1]{\begin{eqnarray}\label{#1}}
\newcommand{\ear}{\end{eqnarray}}
\newcommand{\nn}{\nonumber}

\newcommand{\rf}[1]{(\ref{#1})}
\newcommand{\nl}{ {\hfill \break} }

\newcommand{\ol}{ \overline}

\newcommand{\iso}{ {\cong } }

\newcommand{\Iff}{ {\Leftrightarrow } }

\newcommand{\imp}{\ {\Rightarrow }\ }

\newcommand{\partl}{ {\partial } }
\newcommand{\cl}{ { \mathrm{cl} } }

\newcommand{\R}{ \mbox{\rm I$\!$R} }

\newcommand{\Diff}{ \mbox{\rm Diff} }

\newcommand{\SO}{ \mbox{\rm SO} }

\newcommand{\Int}{ {\mathrm int} }
\newcommand{\Img}{ {\mathrm im} }

\newcommand{\Cinf}{ {\mathcal{C}}^\infty }
\newcommand{\Cyl}{ \mbox{\rm Cyl} }
\newcommand{\Der}{ \mbox{\rm Der}(\Cyl) }

\renewcommand{\setminus}{-}

\def\Journal#1#2#3#4{{#1} {\bf #2}, #3 (#4)}

\def\ATMP{\em Adv. Theor. Math. Phys.}
\def\CMP{\em Commun. Math. Phys.}

\def\GC{\em Grav. Cosmol.}

\def\JMP{\em J. Math. Phys.}

\def\be{\begin{equation}}
\def\ee{\end{equation}}
\def\bea{\begin{eqnarray}}
\def\eea{\end{eqnarray}}

\DeclareFontFamily{U}{rsfs}{}         
\DeclareFontShape{U}{rsfs}{m}{n}{<5> rsfs5 <6><7> rsfs7          %
  <8><9><10><10.95><12><14.4><17.28><20.74><24.88> rsfs10}{}     %
\DeclareMathAlphabet{\mathfs}{U}{rsfs}{m}{n}                     %
\newcommand{\mfs}[1]{\mathfs {#1}}                               %

\newcommand{\sC}{ {\mfs C}}
\newcommand{\sK}{ {\mfs K}}
\newcommand{\sF}{ {\mfs F}}
\newcommand{\sP}{ {\mfs P}}
\newcommand{\sT}{ {\mfs T}}
\newcommand{\sE}{ {\mfs E}}
\newcommand{\sH}{ {\mfs H}}

\newcommand{\olcirc}[1]{\vbox{\mathsurround=0pt                %
  \skip1=0pt plus 1 fil  \skip3=0pt plus 3fil                    %
  \ialign{##\crcr$\hskip\skip3\scriptstyle\circ\hskip\skip1$\crcr%
    \noalign{\kern1pt\nointerlineskip}$\displaystyle{#1}$\crcr}} %
  \vphantom{#1}}                                                 %
\newcommand{\oscirc}[1]{\vbox{\mathsurround=0pt                %
  \skip1=0pt plus 2 fil  \skip3=0pt plus 2fil                    %
  \ialign{##\crcr$\hskip\skip3\scriptstyle\circ\hskip\skip1$\crcr%
    \noalign{\kern1pt\nointerlineskip}$\displaystyle{#1}$\crcr}} %
  \vphantom{#1}}                                                 %

\begin{document}

\centerline{\Large\bf Is Loop Quantum Gravity a QFT ?
\footnote{presented at the Spanish Relativity Meeting, ERE 99, Bilbao}
}
\centerline{\large\bf Martin Rainer}

\begin{center}
Mathematische Physik I, Institut f\"ur Mathematik,  Universit\"at Potsdam,
\\
PF 601553, D-14415 Potsdam, Germany
\\
E-mail: {\em {\tt mrainer@rz.uni-potsdam.de}}
\end{center}

\abstract{
We investigate up to which extend the kinematic
setting of loop quantum gravity can be fit into
a diffeomorphism invariant
setting of algebraic QFT generalizing the Haag-Kastler
setting of Wightman type QFT.
The net of local (Weyl-)algebras resulting from a spin network state
of quantum geometry immediately accommodates isotony and diffeomorphism covariance,
and  formulation of causality becomes possible via of diffeomorphism invariant
foliations of the underlying manifold by cones.
On a spatial horizon, quantum geometry becomes asymptotically
a genuine QFT with infinitely many degrees of freedom,
if the cylinder functions' supporting graphs
intersect the inner boundary spheres in an infinite
number of punctures.
}

\section{Physical motivation of algebraic QFT} 
While classical general relativity usually
employs a Lorentzian spacetime structure,
the most successful approaches for quantum gravity,
such as the canonical quantization of the connection representation,
the loop representation or the spin network representation,
and topological quantum field theory, BF-theories and spin foams
are  invariant under diffeomorphism much more general
than isometries.
The fact that the algebraic structure of quantum gravity should be
background independent motivates a
diffeomorphism invariant generalization of the
Haag-Kastler\cite{HaKa} axiomatic framework of algebraic quantum
(field) theory
to general manifolds without metric structure\cite{Ra9911}.

For a (in a more restricted sense) genuine local QFT
the number of physical degrees of freedom is infinite.
However, the Haag-Kastler framework is a more general setting,
applicable and natural for all local (relativistic)
quantum systems\cite{Ha}. In the case of quantum mechanics
(with a finite number of degrees of freedom)
the localized algebras are just (finite dimensional)
matrix algebras,
in particular algebras which are
von Neumann factors are then of type I.

The general idea of the algebraic approach
is to formulate the theory in terms of local algebras
$\mathcal{A}({\cal O})$
localized on a regions ${\cal O}\subset M$,
rather than in terms of   fields
${\cal O}\ni x\mapsto\phi(x)$ (i.e. operator valued distributions).
The point-like localized fields $\phi$
may typically be tensor fields tensor fields like currents
$J^\mu$ or energy-momentum $T^{\mu\nu}$.
For a typical Q(F)T
the local algebras are concrete
$W^*$ (i.e. von Neumann) algebras,
e.g. given as subalgebras of the algebra of bounded operators
$\mathcal{L}(\mathcal{H})$ on some Hilbert space $\mathcal{H}$
as
\be
{\mathcal{L}}({\mathcal{H}})\supset
{\mathcal{A}}({\cal O}):=
\{\phi(x): x\in {\cal O}\}'' .
\ee
Vice versa, the entity of point-like localized fields can
be reconstructed from localized algebras as
\be
\{ \phi(x) \} =
\cap_{ {\cal O}\ni x } \ol{{\mathcal{A}}({\cal O})} ,
\ee
where the closure $ \ol{\mathcal{A}({\cal O})} $
includes now also unbounded operators giving rise to the distributional
nature of the fields. The properties of the fields
are encoded in the phase space properties of the net of local algebras.
One of the remarkable advantages of the algebraic approach is that it
still works when fields $\phi$ are no longer point-like localized,
but \emph{finitely} localized,
with localization domains of codimension $\geq 1$.
This is e.g. the case for Wilson loops,
cylinder functions, and string or p-brane fields in ${\cal O}$.
In any case the physical information
is given by local observables contained as selfadjoint elements in
local algebras ${\mathcal{A}}({\cal O})$.

Recently the algebraic framework was used successfully
for a proof\cite{Reh} of the holographic hypothesis
(Maldacena conjecture\cite{Mal,Wit}).
The problem there was is the difficulty to find a transformation
between the bulk degrees of freedom
in anti-de Sitter (AdS) space and those of the boundary CFT.
While there apparently does not exist a direct
point transformations between the corresponding quantum fields,
the relation becomes clear for the corresponding
algebras localized in wedge regions on AdS space
and in double cones on its boundary.

A diffeomorphism invariant extension
has  been presented recently\cite{AMMT} for
the axiomatic Osterwalder-Schrader (OS) framework\cite{OS}
of constructive Euclidean QFT\cite{GJ}.
The latter is the axiomatic Euclidean QFT counterpart
of   axiomatic Wightman QFT.
While, the latter is based on vacuum expectations of covariant Wightman
$n$-point functions, the OS approach is based a functional
integral over $n$-point Schwinger functions w.r.t. an invariant measure $\mu$.
Covariance is implemented in the OS setting as the invariance of an
action $S\{f\}:=\int e^{i\phi(f)}d\mu$,
(time-)translation invariance of the vacuum as
(time-)translation invariance of the measure,
locality as OS (time-)reflection positivity,
and uniqueness of the vacuum via an ergodicity (clustering)
property.
Under certain analyticity and regularity assumptions this
correspondence becomes exact, i.e. one can transform
the OS Euclidean field theory in a Wightman QFT
and vice versa. In this case the Euclidean $\phi$ yields
selfadjoint Wightman fields and corresponding local algebras
${\mathcal{A}}({\cal O})=\{e^{i\phi_W(f)^-}\}_{{\mathrm{supp}} f
                                                \subset {\cal O}}$
for a Haag-Kastler setting.

The Osterwalder-Schrader reconstruction
of Euclidean QFT was generalized\cite{AMMT} for diffeomeorphism
invariant theories on manifolds,
essentially replacing Euclidean invariance by diffeomorphism invariance.
A diffeomorphism invariant measure is provided
by a diffeomorphism invariant distribution $P_f(\phi)$.
Complex linear combinations
$\{\sum_i z_i P_{f_i}(\phi)\}=:{\mathcal{A}}$ then yield
a $*$-algebra for the Hilbert space $\mathcal{H}$.
Using a foliation of $M$, reflection positivity
and ergodicity can be formulated as usual. However
(unlike than for standard Euclidean QFT)
strong continuity of the unitary action of time-translations
on  $\mathcal{H}$
has to be added as a separate axiom.
Also gauge-invariance can be taken into account
by a quite natural axiom.
Unfortunately,
there is no obvious
generalization of those analyticity properties which
in the case of the standard Euclidean and Minkowskian QFT
relate the Schwinger functions and Wightman functions.
Without such a generalized "Wick rotation"
it is difficult to relate the OS setting
to local relativistic quantum physics.
While the constructive Osterwalder-Schrader setting provides
an explicit field representation, the more abstract
algebraic framework is more appropriate
for structural investigations, admitting also a more direct
physical interpretation.

Therefore it is the goal of the present investigations
to extend the framework of
algebraic quantum field theory (AQFT) and Haag-Kastler axioms
such that  it becomes also applicable
to theories
which are invariant
under a larger class
of diffeomorphisms, such as quantum gravity.
Within the AQFT setting the question
rises what are the general
classes of diffeomorphisms compatible with a given
algebraic structure.

Earlier attempts\cite{Ra97,RaSa,Ra1} towards a diffeomorphism invariant
algebraic setting for quantum field theory
mainly generalized those axioms that use only
a topological structure on the manifold
already in the usual setting,
such as isotony and  covariance.
However the causality axiom could only be formulated
in a very rudimentary sense, namely by
generalizing Haag duality just on the boundary
of the net. That procedure introduced quite
strange features on the net of $*$-algebras.
In particular it implied the existence of
an Abelian center in the localized algebras
which was then associated with the minimal (interior) boundary\cite{Ra97}.
The existence of such an Abelian center implied
in particular that the algebras could not be
usual CCR Weyl algebras,
since these would have to be simple if the
symplectic form was nondegenerate.

Nevertheless, presently it is known that
the algebraic structure of
a free quantum field theory on Minkowski space
or an asymptotically free field theory on a
(usually globally hyperbolic) curved space-time
is  encoded in a \emph{ causal} net
of C$^*$-algebras.
In particular on Minkowski space,
there exists a strong correspondence
between particular causal sets of Minkowski space
and localized C$^*$-algebras of the net.
There are particular causal sets
which form a topological basis
of Minkowski space,
namely the bounded double cones.

This motivates the present approach where we present
a generalization of this causal net structure
in an a priori background independent
("diffeomorphism invariant") manner.
Here the net has to be background independent,
but still compatible
with a  (metric independent)
notion of causality.
In order to achieve this, we  have to abstract
an appropriate {\em topological} notion
of causality.
Appropriate definitions for such a notion
were given recently\cite{Ra99}.
Those diffeomorphism which preserve
such topological causal structure,
should naturally also leave invariant
the algebraic structure of the net.
A causal topology on a $d+1$-dimensional
manifold then provides a topological notion of
a causal complement on any set.
The next step is then to find a natural implementation
of the correspondence between
causal sets  on a differentiable manifold $M$
and C$^*$-algebras localized on these sets.
This amounts to a causality axiom for the net of  algebras
over $M$.

The distribution of eigenvalues in a concrete measurement
of observables can be extracted with the help of
a corresponding physical state
on the causal net of C$^*$-algebras.
Particularly convenient states for quantum geometry
are the spin-network states.
A state introduces additional structure
which also may serve to distinguish gauge invariances
from more genuine symmetries (like the unitarity of the dynamics).
Given a causal foliation
by spatial slices exterior to a causal horizon
(an interior boundary), those diffeomorphisms which
leave invariant the causal foliation are purely
gauge.

In quantum geometry spin-network states
are given via an embedding of a closed graph into
a particular slice of the foliation.
Then there exists diffeomorphisms which keep
the set of all slices invariant
but change the foliation monotoneously,
preserving the natural order of the slices.
These are topological dilations.
If there was no state two foliations related by such
a change should
be considered as equivalent.
However an embedded graph can eventually
detect a monotonous change in the foliation
by a change in the original relations between
the slices and the graph, which are given by the
intersections of the edges of the graph with
the slices of the foliation.
The relations between an embedded graph
and a foliation are encoded topologically
in the
intersections of edges with slices of a foliations.
Change of this intersection topology by dilating one slice onto
another can result in changes of the C$^*$-algebra.
Therefore dilation diffeomorphisms can not be gauge here,
but rather should correspond to outer (i.e. non-trivially
represented) isomorphisms on the algebras.
\section{Cone causality on  differentiable $d+1$-manifolds} 
Recently, a notion of a causal topology was given\cite{Ra99}
on a general
differentiable $d+1$-manifold $M$ within
any differentiability category.
Let
\beq{scone}
\sC:=\{x\in\R^{d+1}:x_0^2=(x-x_0 e_0)^2\} ,
\sC^+:=\{x\in \sC:x_0\geq 0\} ,
\sC^- :=\{x\in \sC:x_0\leq 0\}
\eeq
be the standard (unbounded double) light cone,
and the forward and backward subcones in $\R^{d+1}$, respectively.
The standard open interior and exterior of $\sC$ is defined as
\beq{scone2}
\sT:=\{x\in\R^{d+1}: x_0^2>(x - x_0 e_0)^2\} ,
\sE:=\{x\in\R^{d+1}: x_0^2<(x - x_0 e_0)^2\} .
\eeq
A \emph{manifold thickening} with thickness $m>0$ is given as
\beq{thickcone}
\sC^m:=\{x\in\R^{d+1}:|x_0^2-(x - x_0 e_0)^2|<m^2\} ,
\eeq
The characteristic topological data of the standard cone
is encoded in the topological relations of all its
manifold subspaces
(which includes in particular also the singular vertex $O$)
and among each other.

A  \emph{(null) cone} at $p\in \Int M$ is defined as
the diffeomorphic image $\sC_p:=\phi_p \sC$
of a diffeomorphism of topological spaces
$\phi_p: \sC \to \sC_p \subset  M $
with $\phi_p(0)=p$, such that
\\
(i) every (differentiable) submanifold $N\subset \sC$
is mapped  diffeomorphically on
a submanifold  $\phi_p(N)\subset M$,
\\
(ii) for any two submanifolds $N_1,N_2\subset \sC$
there exist diffeomorphisms
$ \phi_p(N_1) \cap \phi_p(N_2) \cong N_1 \cap N_2$
and $\phi_p(N_1) \cup \phi_p(N_2)\cong N_1 \cup N_2$,
\\
and (iii) for any two differentiable
curves $c_1,c_2:]-\epsilon,\epsilon[\to\sC$
with $c_1(0)=c_2(0)=p$ it holds
$T_0 c_1=T_0 c_2
\Leftrightarrow
T_p(\phi_p\circ c_1)|_{ ]-\epsilon,\epsilon[ }
=T_p (\phi_p \circ c_2)|_{ ]-\epsilon,\epsilon[ }$.

Condition (iii) says that
the well defined notion of transversality of intersections
at the vertex is preserved by $\phi_p$.
An \emph{(ultralocal) cone structure} on $M$ be
an assignment
$\Int M\ni p\to\sC_p$ of a cone
at every $p\in\Int M$.

Note that, although $\sC_p=\phi_p(\sC)$, $\sT$ and $\sE$ here need
not be diffeomorphic to $\phi_p(\sT)$ and $\phi_p(\sE)$ respectively.
A cone structure on $M$ can in general
be rather wild
with cones at different points totally unrelated
unless we impose a topological connection
between the cones at different points.
The cone structure can be tamed
by the requirement of differentability
of the family $\{p\to\sC_p\}$.
Let a \emph{weak ($\sC$) local cone (LC) structure}   on $ M$ be
a  cone structure which is differentiable
i.e. $\{p\to\sC_p\}$ is a differentiable family.

A weak LC structure at each point $p\in\Int  M$
defines
a characteristic topological space $\sC_p$ of codimension $1$
which is Hausdorff everywhere but at $p$.
In particular $\sC_p$ does not contain any
open $U\ni p$ from the manifold topology of $M$.
This can be  improved by resolving the cone.
Let a  \emph{(manifold) thickened cone} of thickness $m>0$ at $p\in\Int  M$
be the  diffeomorphic image $\sC^{m}_{p}:=\phi_p \sC^m$
of a  diffeomorphism of manifolds
$\phi_p: \sC \to \sC_p \subset  M $
with $\phi_p(0)=p$.
Note that due to the manifold property
a thickened cone is much more simple than a cone itself.
A \emph{thickened cone structure} on $M$ is
an assignment
$\Int M\ni p\mapsto\sC^{m(p)}_{p}$ of a
thickened cone
at every $p\in\Int  M$.

Note that in general the thickness $m$
can vary from point to point in $M$.
However it is natural to require $m:M\to\R_+$
to be differentiable.
A \emph{strong  ($\sC^m$) LC structure}  on $ M$ is
defined as a  differentiable family
of  diffeomorphisms
$\phi_p: \sC^m \to \sC^{m(p)}_{p} \subset M $
with $\phi_p(0)=p$ and
such that the thickness $m$ is a differentiable function on $M$.

Note that a strong LC structure is still much more flexible than a conformal structure.
For any $q\neq p$
the tangent directions given by $T_q\sC_p$ need a priori
not be related
to tangent directions of null curves of $g$,
since the cone (or its thickening) at $p$ is
in general unrelated to that at $q$.

In order to define topologically
timelike, lightlike, and spacelike relations,
and a reasonable causal complement,
we introduce the following causal consistency
conditions on cones.
$M$
is \emph{(locally) cone causal} or \emph{C-causal} in an open region $U$,
if it 
carries a (weak or strong) LC structure
and in  $U$
the following relations between different cones in $\Int M$ hold:
\\
(1) For $p\neq q$ one and only one of the following is true:
\\
(i) $q$ and $p$ are causally \emph{timelike} related, $q\ll p$ $:\Iff$
$ q\in \sF_p$ $\wedge$ $p \in \sP_q $
(or $p \ll  q$)
\\
(ii) $q$ and $p$ are causally \emph{lightlike} related, $q\lhd  p$  $:\Iff$
$ q\in \sC^+_p\setminus \{p\}$ $\wedge$ $p \in \sC^-_q\setminus \{q\} $
(or $p \lhd  q$),
\\
(iii) $q$ and $p$ are causally unrelated,
i.e. relatively \emph{spacelike} to each other, $q\bowtie  p$ $:\Iff$
$ q\in \sE_p$ $\wedge$ $p \in \sE_q $.
\\
(2) Other cases (in particular non symmetric ones)
do not occur.
\\
$M$ is {\it locally C-causal}, if it is C-causal in any  region $U\subset M$.
$M$ is {\it C-causal} if conditions (1) and (2) hold $\forall p\in \sC$.

Let $M$ be C-causal in $U$.
Then, $q \ll  p$ $\Iff$ $\exists r: q\in \sP_r \wedge p\in \sF_r $,
and $q \lhd  p$ $\Iff$ $\exists r: q\in \sC^+_r \wedge p\in \sC^-_r $.

If an open curve $\R\ni s\mapsto c(s)$ or a
closed curve $S^1\ni s\mapsto c(s)$ is embedded in $M$, then in particular
its image is $\Img(c)\equiv c(\R)\cong \R$ or
$\Img(c)\equiv c(S^1)\cong S^1$ respectively,
whence it is free of selfintersections.
Such a curve is called \emph{spacelike}
$:\Iff$ $\forall p\equiv c(s)\in \Img(c) \exists \epsilon :
c|_{]s-\epsilon,s+\epsilon[\setminus\{0\}}\in\sE_{c(s)}$,
and \emph{timelike}
$:\Iff$ $\forall p\equiv c(s)\in \Img(c) \exists \epsilon :
c|_{]s-\epsilon,s+\epsilon[\setminus\{0\}}\in\sT_{c(s)}$.

Note that C-causality of $M$ forbids a multiple refolding intersection
topology for any two cones with different vertices.
In particular along any timelike curve
the future/past cones do not intersect,
because otherwise there would exist points which are simultaneously
timelike and lightlike related.
Continuity then implies that future/past cones in fact foliate the part of $M$
which they cover. Hence, if there exists  a fibration
$\R\hookrightarrow\Int M\twoheadrightarrow \Sigma$,
then C-causality implies in particular that the future/past cones foliate
on any fiber.
Therefore C-causality
allows a reasonable definition of  a causal complement.
A {\em causal complement} in a set $U$ is a map
$P(U)\ni S\mapsto S^{\perp}\in P(U)$ such that
\\
(i) $S \subseteq S^{\perp\perp}$
\\
(ii) $(\bigcup_j S_j)^\perp=\bigcap_j (S_j)^\perp$
\\
(iii) $S\cap S^\perp=\emptyset$

Since C-causality prohibits relative refolding of cones, it also ensures that
$(\sK^q_p)^{\perp\perp}=\sK^q_p$, i.e. double cones are closed under
$(\cdot)^{\perp\perp}$.

A {\em causal disjointness relation} in a net index set $I$
is a symmetric relation $\perp$ such that
\\
(i) $K_1\perp K_0$ $\wedge$ $K_2 < K_1$  $\imp$ $K_2\perp K_0$,
\\
(ii) for any bounded $J\subset I$:
$K_0\perp K$ $\forall K\in J$ $\imp$ $K_0\perp \sup J$,
\\
(iii) $\forall K_1\in I$ $\exists K_2\in I$: $K_1\perp K_2$.
\\
A {\em causal index set} $(I,\perp)$ is a
net index set with a causal disjointness relation $\perp$.

Any conformal class of a Lorentzian metric,
which is globally hyperbolic without any singularities
determines such a causal structure.
In this case the compact open double cones
form a basis of the usual Euclidean $d+1$ topology.
Each open compact double cone $\sK$ is conformally
equivalent to a copy of Minkowski space.
Consider a spatial Cauchy section $\Sigma$ of $M$
and  a  geodesic world line $p:\tau\to M$ intersecting
$\Sigma$ at $p(0)$, where $\tau$ is the proper time
of the observer.
Now for any $\tau>0$ the causal past of $p(\tau)$
intersects $\Sigma$ in an open set $O_{\tau}$.
Then these open sets  are totally ordered
by their nested inclusion in $\Sigma$,
and their order agrees also with the total
order  of  worldline proper time,
\beq{PO}
O_{\tau_1}\subset O_{\tau_2} \Iff \tau_1<\tau_2 .
\eeq

This is the motivation to consider the
partial order related to the flow of time
and the one related to enlargement in space
to be essentially the same, such that
in the absence of an a priori notion of
a metric time, the nested spatial inclusion
will provide a partial order substituting
time. (Of course this is in essence
similar to the old idea in cosmology of time
given by the volume of a closed, expanding
universe.)


Consider now a double cone $\sK$ in $M$
with $O:=\sK\cap\Sigma$ and $\partl O=i^0(\sK)$
and a diffeomorphism $\phi$ in $M$ with pullbacks
$\phi^\Sigma\in\Diff(\Sigma)$ to $\Sigma$
and
$\phi^{\sK}\in\Diff(\Sigma)$ to $\sK$.
If $\phi(\sK)=\sK$, it can be naturally identified
with an element of $\Diff(\sK)$.
($\phi=id_{M\setminus \sK}$ is a sufficient
but not necessary condition for that
to be true.)
If in addition  $\phi(\Sigma)=\Sigma$
then also $\phi(O)=O$, and $\phi|_O$ is a diffeomorphism of $O$.

Let us now consider a $1$-parameter set of double cones $\{\sK_p\}$
sharing $2$ common null curve segments $n_{\pm}\in\partial\sK_p$
from $i^{\pm}(\sK_p)$ respectively to $i\in i^0(\sK_p)$  which they intersect transversally in $\Sigma$.
Let the cones be such that
and parametrized by a line $c$ in $\Sigma$ starting (transversally to  $n$)
at $i$ to some endpoint $f$ on $\partial\Sigma$
(at spatial infinity) such that $p$ is an interior point of $O_p:=\sK_p\cap\Sigma$.
Then we call the limit $W(n_\pm,c):=\lim_{p\to f} K_p$ the wedge
in the surface through $n_\pm$ and $c$.
Note that while in the usual (say Minkowski) metric case a wedge has a quite
rigid structure since $c$ has a canonical location in a surface spanned by
$n_\pm$, its diffeomorphism invariant analogue is less unique in structure.

\section{Axioms for QFT on  manifolds} 
Clearly QFT on a globally hyperbolic space-time manifold
satisfies isotony (N1), covariance (N2), causality (C), additivity (A)
and existence of a (state dependent GNS) vacuum vector (V).
More particular on Minkowski space there
is is a  unique  Poincare-invariant state $\omega$
such that there is a translational
subgroup of isometries with spectrum in the
closure of the forward light cone only.
However there is no reason
to expect such features
in a more general context.
However, an invariant GNS vacuum vector $\Omega$ still exists
for a  globally hyperbolic space-time,
although in general it depends on the choice
of the state $\omega$.
Hence we will now generalize the axioms of AQFT
from globally hyperbolic space-times to differentiable manifolds.

For a given QFT on manifolds, say the example of quantum gravity
examined below, it remains to check which of the
generalized axioms will hold true.

\subsection{General axioms for QFT on a differentiable  manifold} 
\setcounter{equation}{0}
On a differentiable manifold $M$
part of the AQFT structure
can be related to the topological structure of $M$.
The following AQFT axioms are purely topological and
 should hold on arbitrary differentiable manifolds.
Let $M$ be a differentiable manifold  with additional structure $s$ (which may be empty)
and  $\Diff(M,s)$ denote
all diffeomorphisms which preserve $s$.
A $\Diff(M,s)$-invariant algebraic QFT (in the state $\omega$)
can be formulated in terms of axioms on a
net of $*$-algebras ${\cal A}({\cal O})$
(together with a state $\omega$ thereon).
It should at least satisfy
the following axioms:

\nl
{\bf N1 (Isotony):}
\bear{N1}
{\cal O}_1\subset {\cal O}_2 & \imp &
{\cal A}({\cal O}_1)\subset {\cal A}({\cal O}_2)\
\forall {\cal O}_{1,2}\in \Diff(M,s)
\ear
\nl
{\bf N2 (Covariance):}
\bear{N2}
\Diff(M,s)\ni g {\stackrel{\exists}{\mapsto}} U(g) \in U(\Diff(M,s))&:&
{\cal A}(g{\cal O})=U(g){\cal A}({\cal O})U(g)^{-1} \ .
\ear
(N1) and (N2) are purely topological,
involving only the mere definition of the net.
These axioms make sense even without a causal structure
(see also \cite{Ra97}).

If ${\cal A}({\cal O})$ is a $C^*$-algebra with norm $||\cdot||$,
it makes sense to impose the following additional axioms:
\nl
{\bf A (additivity):}
\bear{A}
{\cal O} = \cup_{j} {\cal O}_j &\imp &
{\cal A}({\cal O}) = \cl_{||\cdot||}\left(\cup_{j} {\cal A}({\cal O}_j)\right) \ .
\ear
\nl
{\bf V (Invariant Vacuum Vector):}
Given a state $\omega$, there exists
a representation $\pi_\omega$
on a Hilbert space ${\cal H}_\omega$
such that
\bear{V}
\exists \Omega\in{\cal H}_\omega, ||\Omega||=1 &:&
\nn\\
\mbox{\rm (cyclic)}&&
\left(\cup_{\cal O} {\cal R}({\cal O})\right) \Omega
{\stackrel{\rm dense}{\subset}} {\cal H}_\omega
\nn\\
\mbox{\rm (invariant)}&&
U(g)\Omega=\Omega
\ , \quad
g\in \Diff(M,s)\ .
\ear
Note: For any $*$-algebra, the representation $\pi_\omega$ is given by
the GNS construction, ${\cal H}_\omega$ is the GNS Hilbert space.
Properties of $\Omega$ are induced by corresponding properties
of the state $\omega$.
The main issue to check is the invariance under a
unitary representation $U$ of \Diff(M,s).

\subsection{Axioms for QFT on a manifold with cone causality} 
\setcounter{equation}{0}
With a notion of causality on a differentiable manifold $M$
as defined in the previous section,
the algebraic structure of a QFT
should be related to the causal differential structure of $M$
by further axioms abstracted from
the space-time case.
In this case it is natural to consider nets of
von Neumann algebras.
On a causal differential manifold $M$ (in the sense defined above)
the algebraic structure of a QFT
should satisfy the following axioms which
require the notion of a causal complement.
Let $M$ be a causal differentiable manifold
with additional structure $s$ (which may be empty)
and  $\Diff(M,s)$ denote
all differentiable diffeomorphisms which preserve $s$,
where $s$ is at least a causal structure,
eventually with some additional structure $s'$.
A $\Diff(M,s)$-invariant algebraic QFT in the state $\omega$
is a net of von Neumann-algebras
${\cal R}({\cal O}$
with a state $\omega$
satisfying the following axioms:
\nl
{\bf C (causality):}
\bear{C}
{\cal O}_1\perp {\cal O}_2 &\imp&
{\cal R}({\cal O}_1)\subset {\cal R}({\cal O}_2)' \ .
\ear
\nl
{\bf CA (causal additivity):}
\bear{CA}
{\cal O} = \cup_{j} {\cal O}_j &\imp &
{\cal R}({\cal O}) = \left(\cup_{j} {\cal R}({\cal O}_j)\right)'' \ .
\ear
Remarks:
In the case that the net has
both inner and exterior boundary,
\rf{C} had been  weekened in \cite{Ra97}
to a generalization of Haag duality on the boundary of the net.
Here we do not assume a priori the existence of such a
boundary of the net. However an example
of quantum geometry with such a boundary structure
is discussed below.

Given  a net of $C^*$ algebras consistent
with a norm $||\cdot||$, it made sense
to impose (A) above.
If the algebras are in particular also von Neumann ones
(A) should be sharpened to (CA).
In the general case of $*$-algebras (not necessarily $C^*$ ones)
the algebraic closure has no natural topological analogue,
and hence there is
no obvious definition of additivity.
Therefore in \cite{Ra97} neither (A) nor (CA) was assumed.

\section{Loop quantum gravity exterior to $\sH$} 
\setcounter{equation}{0}
The intersection of a spatial slice with
a causal horizon $\sH$ is a $d-1$-dimensional sphere $S^{d-1}$.
The latter may be viewed as the boundary of a
minimal $d$-dimensional open set ${O}_{\rm min}$
contained in any
larger $d$-dimensional open set  ${O}_{\rm max}$
within a spatial slice
which extends through all of the region exterior
of the horizon up to spatial infinity $i^0$.
In \cite{Ra97} a generalization of Haag duality
was implemented algebraically,
by demanding that the commutant of the (asymptotic) global
algebra equals a minimal Abelian center algebra
located over the minimal set.

On a region causally exterior to $\sH$,
on any $d$-dimensional spatial slice $\Sigma$,
there exists a net of Weyl algebras
for states with
a  countable, {\em infinite} number of intersection
points of edges and
transversal $(d-1)$-faces
within any neighbourhood of the spatial boundary
$\sH \cap\Sigma\iso S^{d-1}$.

$\Sigma$ be a spatial slice.
C-causality constrains the algebras
localized within $\Sigma$.
On $\Sigma$ it should hold
\begin{equation}\label{discom}
{\cal O}_1\cap {\cal O}_2=\emptyset
\quad \imp \quad [{\cal A}({\cal O}_1), {\cal A}({\cal O}_2)]=0 .
\end{equation}

A (spin network) state $\omega$ over the algebra ${\cal A}(\Sigma)$
may be defined by representations on
a closed, oriented differentiable graph $\gamma$
embedded in $\Sigma$,
with an infinite  number of differentiable edges $e\in E$
intersected transversally by a
differentiable $d-1$-dimensional oriented surface $S$
at a countable of intersection vertices $v\in V$.
Let $C_{\gamma}\in \Cyl$ be a $\Cinf$ Cylinder function with respect to an
$\SO(d)$ holonomy group  on $\gamma$, i.e.
on each closed {\em finite} subgraph $\gamma'\subset\gamma$
it is $C_{\gamma'}:=c(g_1,\ldots,g_N)$ where $g_k\in\SO(d)$,
and $c$ is a differentiable function.
With test function $f$ a derivation
$X_{S,f}$ on $\Cyl$ is defined by
\beq{deriv}
 X_{S,f}\cdot C_{\gamma}:=\frac{1}{2}
 \sum_{v\in V}\sum_{e_v\in E:\partl e_v\ni v}
 \kappa(e_v) f^i(v) X^i_{e_v} \cdot c ,
\eeq
where $\kappa(e_v)=\pm1$ above/below $S$
(for the following purposes we may just exclude the tangential
case $\kappa(e_v)=0$)
and $ X^i_{e_v} \cdot c $ is the action of the left/right invariant
vector field (i.e. $e_v$ is oriented away from/towards the
surface $S$) on the argument of $c$ which corresponds to the
edge $e_v$. Let $\Der$ denote the span of such derivations.

Here the classical (extended) phase space is the cotangent
bundle  $\Gamma=T^*{\mathcal{C}}$
over a space $\mathcal{A}$ of (suitably regular) finitely localized
connections.
Let $\delta = ( {\delta_A} , {\delta_E} ) \in T_{e}\Gamma$.
With suitable boundary conditions,
a (weakly non-degenerate) symplectic form $\Omega$ over $\Gamma$
acts via
\bear{sympl}
\Omega|_{(A_e,E_e)}
\left ( \delta, \delta' \right )
&:=&
{1 \over \ell^2} \int_{\Sigma} {\rm Tr}
\left [  *E \wedge A'  -   *E' \wedge A  \right ] .
\ear
After lifting from $\mathcal{C}$ to $\Gamma$,
the cylinder functions $q\in\Cyl$ serve as
(gauge invariant) classical elementary
configuration functions on $\Gamma$.
The derivations $p\in\Der$ serve as classical elementary momentum
functions on $\Gamma$. They are obtained as the Poisson-Lie action
of $2$-dimensionally smeared duals of densitized triads $E$.
$\Cyl\times\Der$ has a Poisson-Lie structure
\beq{Poisson}
\{(q,p),(q',p'))\}:=(p q'- p' q  ,[p,p']) ,
\eeq
where $[p,p']$ denotes the Lie bracket of $p$ and $p'$.
An antisymmetric bilinear form on $\Cyl\times\Der$ is given by
\bear{pres}
\Omega_{0}
\left ( {(\delta_q,\delta_p)}, {(\delta_q',\delta_p')} \right )
&:=&
\int_{{\mathcal{C}_{\gamma\cup\gamma'}}/
{\mathcal{G}_{\gamma\cup\gamma'}}}d\mu_{\gamma\cup\gamma'}
\left [  p  q'  -   p' q  \right ] ,
\ear
where $q,q'\in \Cyl$ have support on $\gamma$ resp. $\gamma'$,
with $p  q' -  p' q\in \Cyl$
integrable over ${\mathcal{C}_{\gamma\cup\gamma'}}/
{\mathcal{G}_{\gamma\cup\gamma'}}$ with
measure $d\mu_{\gamma\cup\gamma'}$.

On $T_e\Gamma$, the symplectic form $\Omega$  yields functions
of the form $\Omega((\delta_A,\delta_E),\cdot)$.
Canonical quantization then associates
to any function $\Omega(f,\cdot)$
a selfadjoint operator $\hat\Omega(f,\cdot)$
and a corresponding
unitary Weyl element $W(f):=e^{i\hat\Omega(f,\cdot)}$,
both on some extended Hilbert space.
With multiplication
$W(f_1) W(f_2) := e^{i\Omega(f_1,f_2)}W(f_1+f_2)$,
and conjugation $*: W(f)\mapsto W(-f)$,
the Weyl elements generate a $*$-algebra.
A norm on $\Gamma$ is defined by
$\|f\|:=\frac{1}{4}\sup_{g\neq 0}\frac{\Omega(f,g)}{<g,g>}$.
The $C^*$-closure under the sup-norm then generates
a  $C^*$-algebra $CCR(W(f),\Omega)$.
With regular $\Omega$ this CCR Weyl algebra is simple,
i.e. there is no ideal.
Observables of quantum $3$-geometry are
then the selfadjoint elements within a
gauge and $3$-diffeomorphism invariant $C^*$-subalgebra
${\cal A}_{\gamma}\subset C^*(W(f),f\in\Gamma)$.
In a gauge and $3$-diffeomorphism invariant
representation of ${\cal A}_{\gamma}$,
typical observables in are represented by
configuration multiplication operators $C_{\gamma}\in \Cyl$
on Hilbert space $\mathcal{H}_{\gamma}$,
and by gauge-invariant and 3-diffeomorphism invariant
combinations of derivative operators $X_{S,f}\in \Der$,
like e.g. a certain quadratic combination which yields
the area operator.
%

For each finite $\gamma'\subset\gamma$, the sets $E(\gamma')$ and $V(\gamma')$
of edges resp. vertices of $\gamma'$ are finite. Then the
connections
${\mathcal{C}}_{\gamma'}=\prod_{e\in E(\gamma')}G_e\cong G^{E(\gamma')}$
and the gauge group
${\mathcal{G}}_{\gamma'}=\prod_{v\in V(\gamma')}G_v\cong G^{V(\gamma')}$
on $\gamma'$ inherit a unique measure from the measure on $G$
(for compact $G$ the Haar measure).
The action of $\mathcal{G}_{\gamma'}$ on $\mathcal{C}_{\gamma'}$ is defined by
$(g A)_e := g_{t(e)} A_e g_{s(e)}^{-1}$ where $s$ and $t$ are the source and target
functions $E(\gamma')\to V(\gamma')$ respectively.
This action gives rise to gauge orbits and a corresponding projection
$\mathcal{C}_{\gamma'}\twoheadrightarrow \mathcal{C}_{\gamma'}/\mathcal{G}_{\gamma'}$.
The projection induces the measure on $C_{\gamma'}/\mathcal{G}_{\gamma'}$.
Bounded functions w.r.t. to that measure define then
the gauge invariant Hilbert space
$\mathcal{H}_{\gamma'}:=L(\mathcal{C}_{\gamma'}/\mathcal{G}_{\gamma'})$.

However, over finite graphs, all is still QM rather than QFT.
In order to obtain
an infinite number of degrees of freedom
on any finite localization domain which includes
the inner boundary $S^{d-1}$ (the intersection $S^{d-1}$
of $\sH$ and $\Sigma$),
let  $S^{d-1}$ be intersected
by an infinite number of edges of some graph $\gamma$
in the exterior spatial neighborhood of  $\sH$.
In the $3+1$-dimensional case, evaluation of the area operator
on the puncture of the boundary $S^2$ from edge $p$
yields a quantum of area proportional to
$j_p(j_p+1)$ for edge $p$ carrying a spin-$j_p$
representation of the group $G$.
Since $S^2$ is compact the punctures should have at least one
accumulation point.
Hence for typical configurations in the principal representation,
near that accumulation point the area will explode to infinity.
When almost all punctures are located in arbitrary small neighborhoods
of a finite number $n$ of accumulation points,
corresponding states represent quantum geometries of a black hole with $n$
stringy hairs extending out to infinity. In particular, the $n=1$ case was discussed
in more detail in \cite{MR99bh}.
\section{Conclusion} 
Above we indicated how loop quantum gravity
can be fit into
a generalization of the
Haag-Kastler axiomatic framework of algebraic Q(F)T
on differentiable manifolds.

For quantum geometry, the configuration fields
are certain cylinder functions with support finitely localized
on a graph within in an open set of the index set
of a Haag-Kastler net over a differentiable manifold.
Such a net is given by certain $C^*$-algebras (for loop quantum gravity:
Weyl algebras) localized on certain open sets of the manifold
in a background independent manner
(in particular without recurrence to a metric),
via basic algebraic topological
key properties like isotony, and diffeomorphism covariance.
A state-dependent  diffeomorphism invariant GNS vacuum always exists.
(If for a net of von Neumann algebras,
this GNS representation is cyclic and separating,
the modular group can be extracted.)

The central axiom of the Haag-Kastler framework is
causality.
Previous attempts\cite{Ra97} to implement its analogue
in the more general general diffeomorphism invariant context,
as a duality on the boundary of the net, led to
very unusual consequences for the net of local algebras
and the encoded QFT.
Above we therefore proposed another approach,
via diffeomorphism invariant foliations of $M$
by cones\cite{Ra99}.
On a general differentiable manifold
of dimension $d+1>2$, causality is obtained from
a $d+1$-parameter family of cones, where each cone is homeomorphic
to a $d$-dimensional standard cone.
With this notion of causality on the manifold,
the causality axiom for the net of algebras can be
formulated in analogy to the case of QFT in metric backgrounds.
Moreover, locally the causal structure would fix already
a conformal structure, provided it exists.
For minisuperspace geometries, this supports
a former proposal of conformal
covariant quantization\cite{MR95}.
In general, it rises the question, up to which extend a causal,
diffeomorphism
invariant QFT implies the existence of a conformal background.

States of a genuine QFT should reflect infinitely many degrees
of freedom. If there is an infinite number of punctures
on the inner  boundary
$S^{d-1}$, resulting from intersections
with the supporting  graph of the exterior quantum geometry,
then loop quantum gravity near that horizon becomes
a genuine QFT with an infinite number
of degrees of freedom.
(Note that the generalization of spin networks
to infinite graphs requires nevertheless a particular care with
the reduction of infinite tensor products.)

The resulting picture seems to be consistent with the holographic
principle, by which a conformal QFT arises on an appropriate
boundary screen (e.g. a horizon).

The present causal approach gives also progress towards
a natural formulation of the additivity axiom over double cones.
The diffeomorphism invariant analogue of spectrum positivity property,
the modular structure and the geometric interpretation
of the net, the type of the algebras involved,
and the particular role of dilations are some more questions
which require further investigations.


\begin{thebibliography}{99}

\bibitem{HaKa}
R. Haag and D. Kastler, \Journal{\JMP}{5}{848}{1964}.

\bibitem{Ra9911}
M. Rainer,
Algebraic Quantum Theory on Manifolds:
A Haag-Kastler Setting for Quantum Geometry,
gr-qc/9911076.
%


\bibitem{Ha}
R. Haag, Local Quantum Physics (Springer Verlag, Berlin, 1992).

\bibitem{Reh}
{
K.-H. Rehren
},
{
Algebraic Holography
},
{
}
http://xxx.lanl.gov/abs/hep-th/9905179

\bibitem{Mal}
{
J. M. Maldacena
},
{
The Large N Limit of Superconformal Field Theories and Supergravity
},
\Journal{\ATMP}{2}{231-252}{1998}.

\bibitem{Wit}
{
E. Witten
},
{
Anti de Sitter Space and Holography
},
\Journal{\ATMP}{2}{253-291}{1998}.

\bibitem{AMMT}
{
A. Ashtekar, D. Marolf, J. Mourão, and T. Thiemann
},
{
Osterwalder-Schrader Reconstruction and Diffeomorphism Invariance
}
{
}
quant-ph/9904094

\bibitem{OS}
K. Osterwalder and R. Schrader,
Axioms for Euclidean Green's functions.
I,
\Journal{\CMP}{31}{83-112}{1973};
II,
\Journal{\CMP}{42}{281-305}{1975}.

\bibitem{GJ}
J. Glimm and A. Jaffe,
{Quantum Physics, 2nd. ed.},
(Springer, New York, 1987).





\bibitem{Ra97}
M. Rainer,
The Role of Dilations in Diffeomorphism Covariant Algebraic Quantum Field Theory,
 gr-qc/9710081.

\bibitem{RaSa}
M. Rainer and H. Salehi,
A Regularizing Commutant Duality for a Kinematically Covariant
Partial Ordered Net of Observables,
gr-qc/9708059.

\bibitem{Ra1}
M. Rainer,
General Regularized Algebraic Nets for
General Covariant QFT on Differentiable Manifolds,
gr-qc/9705084.
%



\bibitem{Ra99}
M. Rainer,
Cones and Causal Structures on Topological and Differentiable Manifolds,
gr-qc/9905106, to appear in \Journal{\JMP}{40}{}{1999}.





\bibitem{MR99bh}
M. Rainer,
On the fundamental length of quantum geometry and the black hole entropy,
gr-qc/9903091.

\bibitem{MR95}
M. Rainer,
\Journal{\GC}{1}{121-130}{1995}.
\end{thebibliography}
\end{document}